\documentclass[twocolumn,english,pra,showpacs]{revtex4-1}
\usepackage[T1]{fontenc}
\usepackage[latin9]{inputenc}
\usepackage{amsmath}
\usepackage{graphicx}
\usepackage{amssymb}
\usepackage{esint}

\makeatletter
\@ifundefined{textcolor}{}
{%
 \definecolor{BLACK}{gray}{0}
 \definecolor{WHITE}{gray}{1}
 \definecolor{RED}{rgb}{1,0,0}
 \definecolor{GREEN}{rgb}{0,1,0}
 \definecolor{BLUE}{rgb}{0,0,1}
 \definecolor{CYAN}{cmyk}{1,0,0,0}
 \definecolor{MAGENTA}{cmyk}{0,1,0,0}
 \definecolor{YELLOW}{cmyk}{0,0,1,0}
 }

\makeatletter
\usepackage{dcolumn}
\usepackage{bm}
\usepackage{amsthm}\@ifundefined{definecolor}
 {\@ifundefined{definecolor}
 {\usepackage{color}}{}
}{}

\makeatother

\usepackage{babel}

\makeatother

\usepackage{babel}

\begin{document}
\newtheorem{conjecture}{Conjecture}\newtheorem{corollary}{Corollary}\newtheorem{theorem}{Theorem}
\newtheorem{lemma}{Lemma}
\newtheorem{observation}{Observation}
\newtheorem{definition}{Definition}\newtheorem{remark}{Remark}\global\long\global\long\def\ket#1{|#1 \rangle}
 \global\long\global\long\def\bra#1{\langle#1|}
 \global\long\global\long\def\proj#1{\ket{#1}\bra{#1}}

\title{The amplification of weak measurements under quantum noise}

\author{Xuanmin Zhu{$^{1}$}}\email{zhuxuanmin2006@163.com}  \author{Yu-Xiang Zhang$^{2}$}\email{iyxz@mail.ustc.edu.cn}

 \affiliation{$^{1}$ School of Physics and Optoelectronic Engineering, Xidian University, Xi'an 710071, China \\
 $^{2}$Hefei National Laboratory for Physical Sciences at Microscale, University of Science and Technology
 of China, Hefei, Anhui 230026, China }

\date{\today}

\pacs{03.67.-a, 03.65.Ta}
\begin{abstract}
The influence of outside quantum noises on the amplification of weak measurements is investigated. Three typical quantum noises are discussed. The maximum values of the pointer's shifts decrease sharply with the strength of the depolarizing channel and phase damping. In order to obtain significant amplified signals, the preselection quantum systems must be kept away from the two quantum noises. Interestingly, the amplification effect is immune to the amplitude damping noise.
\end{abstract}
\maketitle
\section{Introduction}
In the quantum weak measurements introduced by Aharanov \emph{et al.}~\cite{aav}, when the coupling strength between the measuring device and the quantum systems is very weak, the values of the meter's readings could be much larger than those obtained in the traditional quantum measurement. Because of such amplification effect, the weak measurement strategies have been used to implement a lot of high-precise measurements, such as observing the tiny spin Hall effect of light~\cite{hall}, and amplifying small transverse deflections and frequency changes of optical beams~\cite{def1}. Besides the advantages in measuring small signals, weak measurements are very useful for exploring the fundamental problems of quantum mechanics~\cite{hardy1,hardy2,exp2,Lundeen,Kocsis,Cho}. Since its importance in applications and fundamental theories, weak measurement has attracted much attention~\cite{w1,w01,w2,w3,w4,w5,w6,w06,w7,w8,w9,w10,w11,w12,w13,w14,w15,w16,w17,w18,w19}.

For the random errors which can be reduced by repeated measurements~\cite{e1}, there is also an open question whether the weak measurements can improve the signal-to-noise ratio (SNR) significantly compared with the ordinary measurements~\cite{d1,d2,d3,d4,d5}. But for the system errors which cannot be reduced by repeated measurements~\cite{e1}, the amplification effect of weak measurements can effectively reduce the relative system errors. This is the reason why a lot of super-precise measurements are accomplished with the methods of weak measurement. From the results in the literature~\cite{zhu,wu,ag,susa,pang}, it is concluded that when the coupling strength is weak, and the measure device state is the typical Gaussian wave function, the maximal shifts of the pointer momentum and position are are respectively equal to $\Delta p$ (the standard deviation of the momentum of the pointer initial state) and $\Delta q$ (the standard deviation of the position of the pointer initial state) which are independent with the coupling strength $g$.

For the fixed measure device and coupling strength, the maximal shifts of the pointer are obtained by choosing nearly orthogonal preselection and postselection states (PPS). Compared with classical states, quantum states are much more sensitive to the disturbance of the outside environment. If the quantum states suffers from some noise, could we obtain the significant amplification as we expected? If not, what the amplification limit will be? In this article, we will study these two questions. As the postselection states are realized by a sequential strong measurement after the weak interaction between the device and systems, and what we need are the outcomes. So we only study the maximum shifts of the pointers when the preselection states are disturbed by the noise from the outside environment.

This paper is organized as follows. In Sec. II we will study the relationships between the modulus of the preseletion states and the amplification. The maximum
pointer shifts are given under the quantum noise for the Gaussian and qubit meters in Sec. III. A short conclusion
is presented in Sec. IV.

\section{The modulus of the preselection states and the amplification}
In weak measurements, the quantum systems that to be measured are prepared in some state $|\psi_{i}\rangle_s$. During the storage and transmission processes, the quantum systems are always correlated with the outside environment unavoidably. And the pure preslection state $|\psi_{i}\rangle_s$ may change into a mixed state $\rho_{s}$. In this article, it is assumed that the preselection state has been changed into a mixed state $\rho_{s}$ before entering measure device. The weak interaction between the quantum systems and measure device could be described by an impulse Hamiltonian
\begin{equation}\label{e1}
\mathbf{H}=-i g\delta(t-t_0)\mathbf{A}\otimes \mathbf{D},
\end{equation}
where $g$ is the coupling strength with $g\geq 0$, $\mathbf{A}$ is the observable operator of the quantum systems, and $\mathbf{D}$ is the observable operator of measure  device. The measure device state is denoted by $\rho_{d}$, after the interaction, the combined state of the system and the measure device state is
\begin{equation}\label{e2}
\rho'_{sd}=\mathbf{U}\rho_{s}\otimes\rho_{d}\mathbf{U}^{\dagger},
\end{equation}
where $\mathbf{U}=\exp(ig\mathbf{A}\otimes\mathbf{D})$ with $\hbar=1$. Different with ordinary measurements, weak measurements require a postselection to select the system into state $|\psi_f\rangle_s$. After the postselection, with the notation $\mathbf{\Pi}_{f}=|\psi_f\rangle_s\langle\psi_f|$, the final measure device state is
\begin{equation}\label{e3}
\rho'_{d}=\frac{\text{tr}_{s}\left(\mathbf{\Pi}_{f}\otimes\mathbf{I}_{d}\rho'_{sd}\right)}{\text{tr}_{sd}\left(\mathbf{\Pi}_{f}\otimes\mathbf{I}_{d}\rho'_{sd}\right)}.
\end{equation}
In weak measurements, the readings are the expectation values of the pointer conditioned on obtaining the postselectin state $\mathbf{\Pi}_f$. If the observable operator of the pointer to be recorded is $\mathbf{R}$, the shift of $\mathbf{R}$ is
\begin{equation}\label{e4}
\delta R'=\text{tr}(\mathbf{R}\rho'_{d})-\text{tr}(\mathbf{R}\rho_{d}).
\end{equation}
In order to search for the amplification effects, we should choose appropriate PPS ($\rho_{s}$ and $\mathbf{\Pi}_{f}$) to obtain the maximum value of $|\delta R'|$.

In quantum information, the most important and useful systems are qubits, and most of the weak value amplification experiments are based on qubits. In this article, the quantum systems to be measured are qubits. And two types of measure devices are used to analyze the amplification affected by the quantum noise. One is the continuous Gaussian type and the other is the discrete qubit meter.
Generally, a qubit state $\rho_{s}$ could be represented as a vector in the Bloch sphere
\begin{equation}\label{e5}
\rho_{s}=\frac{\mathbf{I}+\vec{r} \cdot \vec{\sigma}}{2},
\end{equation}
where $\vec{r}=(r_x,r_y,r_z)$ is a vector in the Bloch sphere with modulus $r=\sqrt{r_x^2+r_y^2+r_z^2}\in[0,1]$, and the components of $\vec{\sigma}=(\sigma_x,\sigma_y,\sigma_z)$ are the Pauli matrices. The qubit state $\rho_{s}$ can also be expressed as
\begin{equation}\label{e6}
\rho_{s}=\frac{(1-r)\mathbf{I}}{2}+r|\psi_i\rangle_{s}\langle \psi_i|,
\end{equation}
where $|\psi_i\rangle_{s}$ is a normalized pure qubit state.
\subsection{The Gaussian type measure device}
The measure device discussed in this subsection is the Gaussian type. Without loss generality, the Gaussian wave function is assumed centered on $q=0$ and $p=0$. In the position and momentum representations, the wave function is respectively written as
\begin{equation}\begin{split}\label{e7}
&\Phi(q)=\frac{1}{ (2 \pi\Delta^{2})^{\frac{1}{4}}}\exp({-\frac{q^2}{4\Delta^2}}),\\
&\Phi(p)=\frac{(2\Delta^{2} )^{\frac{1}{4}}}{\pi^{\frac{1}{4}}}\exp({-\Delta^2 p^2}),
\end{split}\end{equation}
where $q$ and $p$ are the position and momentum variables with the standard deviations $\Delta q= \Delta$ and $\Delta p= \frac{1}{2\Delta}$. Suppose the observable of the qubit to be measured is $\mathbf{A}=\vec{\sigma}\cdot \vec{n}$, where $\vec{n}=(n_x,n_y,n_z)$ is a unit vector, the operator of the measure device $\mathbf{D}=q$. From Eq. (\ref{e1}), the impulse interaction Hamiltonian could be written as
\begin{equation}\label{e8}
\mathbf{H}=-ig\delta(t-t_0) \vec{\sigma}\cdot \vec{n}\otimes q.
\end{equation}

Denoting the set of the eigenstates of $\vec{\sigma}\cdot \vec{n}$ as the basis $\{|0\rangle_s,|1\rangle_s\}$, the pure state $|\psi_i\rangle_s$ in Eq. (\ref{e6}) could be expressed as $|\psi_i\rangle_s=\alpha_1|0\rangle_s+\beta_1|1\rangle_s$, and the postselection state could be expressed as $|\psi_f\rangle_s=\alpha_2|0\rangle_s+\beta_2|1\rangle_s$. From Eqs. (\ref{e2}), ({\ref{e3}}), ({\ref{e6}}), ({\ref{e7}}) and (\ref{e8}), we get the final measure device state
\begin{equation}\label{e9}
\rho'_{D}=\frac{B_1+B_2+B_3+B_4}{Pro},
\end{equation}
where  $B_1=((1-r|\alpha_2|^2)/{2}+r|\alpha_1|^2|\alpha_2|^2)e^{igq}\rho_{D}e^{-igq}$, $B_2=((1-r|\beta_2|^2)/{2}+r|\beta_1|^2|\beta_2|^2)e^{-igq}\rho_{D}e^{igq}$,
$B_3=r(\alpha_1^*\beta_1\alpha_2\beta_2^*)e^{-igq}\rho_{D}e^{-igq}$, $B_4=r(\alpha_1\beta_1^*\alpha_2^*\beta_2)e^{igq}\rho_{D}e^{igq}$, $\rho_{d}={1}/{\sqrt{2\pi\Delta^2}}\iint e^{-q_1^2/{4\Delta^2}}e^{-q_2^2/{4\Delta^2}}|q_1\rangle\langle q_2|\text{d}q_1\text{d}q_2$ given by Eq. (\ref{e7}), and $Pro=(1-r)/2+r(|\alpha_1|^2|\alpha_2|^2+|\beta_1|^2|\beta_2|^2+2\mathbf{Re}(\alpha_1^*\beta_1\alpha_2\beta_2^*)e^{-2\Delta^2g^2})$ is the probability of obtaining the postselection state.

Using Eq. (\ref{e4}), the average shifts of the pointer momentum and position are
\begin{equation}\begin{split}\label{e10}
\delta p'&=\frac{g((1-r)(|\alpha_2|^2-|\beta_2|^2)+2r(|\alpha_1|^2|\alpha_2|^2-|\beta_1|^2|\beta_2|^2))}{2Pro}, \\
\delta q'&=\frac{4rg\Delta^2e^{-2g^2\Delta^2}\mathbf{Im}(\alpha_1^*\beta_1\alpha_2\beta_2^*)}{Pro}.
\end{split}\end{equation}
By derivation, we obtain the maximum shifts
\begin{equation}\begin{split}\label{e11}
|\delta p'|_{max}=\frac{g}{\sqrt{1-r^2e^{-4\Delta^2g^2}}},\\
|\delta q'|_{max}=\frac{2rg\Delta^2e^{-2\Delta^2g^2}}{\sqrt{1-r^2e^{-4\Delta^2g^2}}}.
\end{split}\end{equation}
The detail derivation and the conditions of attaining the maximum shifts are given in Appendix A. The relationships between the maximum shifts and the moduli of the preselection states are pictured in Fig. 1.

From Fig. 1, it can be seen that the maximum shifts decrease sharply with the modulus $r$. In order to obtain significant amplification effects of weak measurements, the modulus of the qubits should be close to 1. In other words, the quantum states must be kept in pure states for searching for markedly amplified shifts.

\begin{figure}[t]
\centering \includegraphics[scale=0.5]{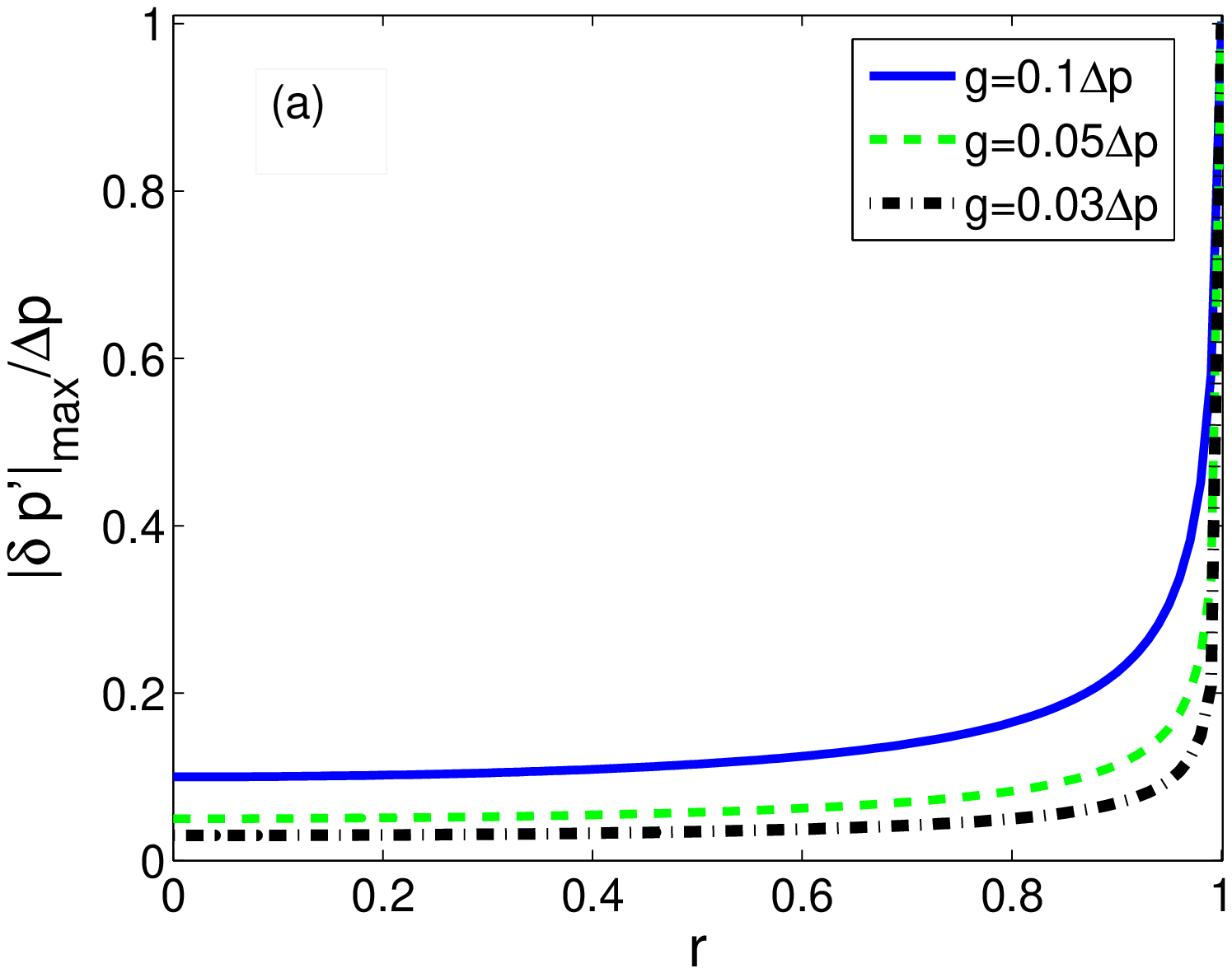} \includegraphics[scale=0.5]{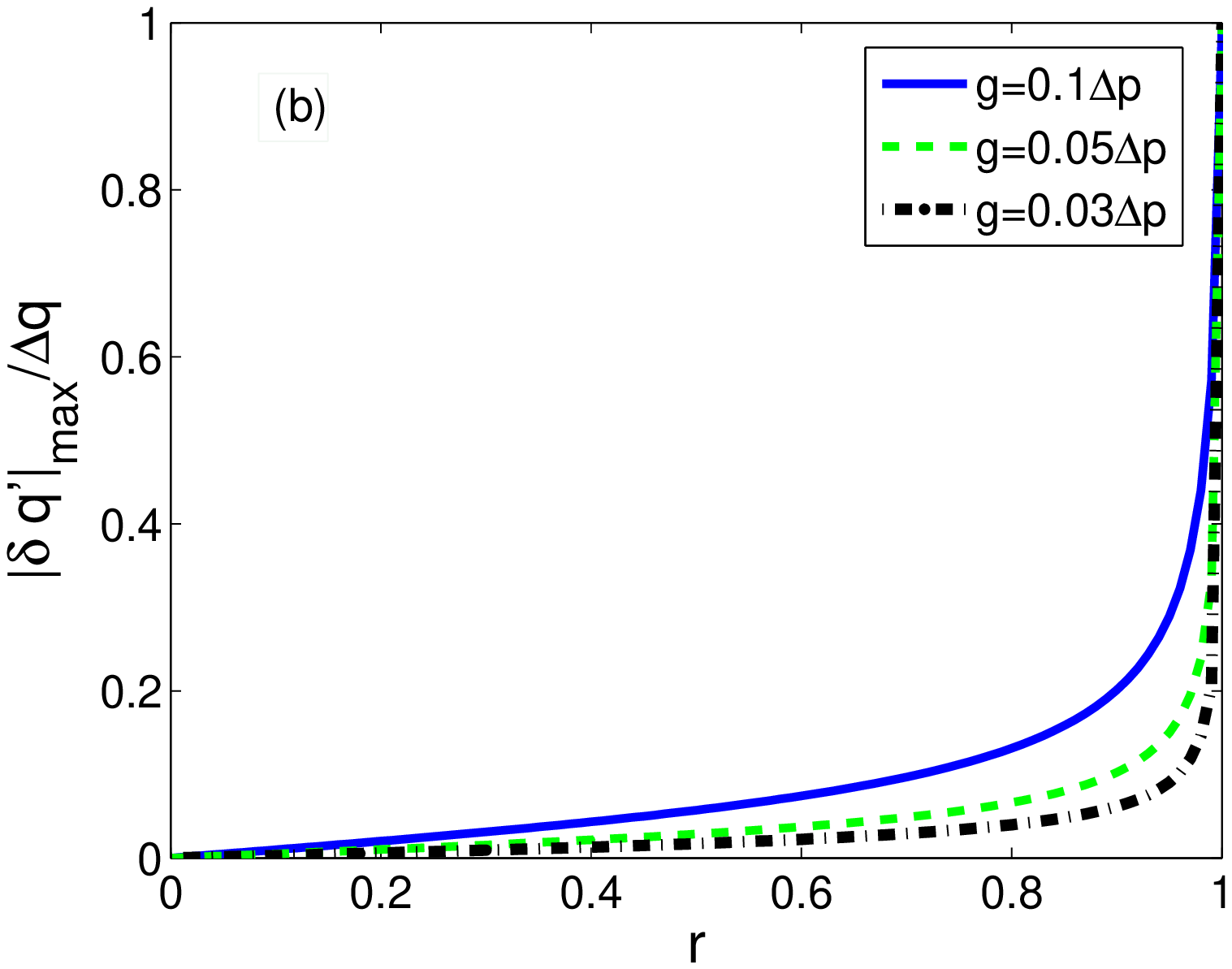} \caption{(Color online) The relationship between the maximum shifts
and the modulus $r$ of the preselection state with $g=0.1\Delta p$, $g=0.05\Delta p$, and $g=0.03\Delta p$.}
\label{fig:1}
\end{figure}
\subsection{The qubit type measure device}
In this subsection, the qubit type measure device is used to study the maximum readings. Since the qubit type measure device is not widely discussed in theory, first, we will show how to obtain amplified readings using qubit measure device. The initial state of the qubit measure device is pure state denoted as
\begin{equation}\label{e12}
|\phi_i\rangle_{d}=|0\rangle_{d},
\end{equation}
and its expectation value of the observable $\mathbf{O}=|1\rangle_{d}\langle 1|$ is 0, where $_d\langle0|1\rangle_{d}=0$. The initial state of the systems to be measured is denoted by $|\psi_i\rangle_{s}=\alpha_{1}|0\rangle_{s}+\beta_{1}|1\rangle_{s}$ with $|\alpha_1|^2+|\beta_1|^2=1$. The impulse weak interaction between the device and system is expressed in the Hamiltonian
\begin{equation}\label{e13}
\mathbf{H}=-ig\delta(t-t_0)\sigma_{z,s}\otimes\sigma_{x,d},
\end{equation}
where $\sigma_{z,s}$ and $\sigma_{x,d}$ are Pauli matrices of the systems and the measure device respectively. The device operator $\sigma_{x,d}=|+\rangle_{d}\langle +|-|-\rangle_{d}\langle-|$, where $|+\rangle_{d}=1/{\sqrt{2}}(|0\rangle_{d}+|1\rangle_{d})$ and $|-\rangle_{d}=1/\sqrt{2}(|0\rangle_{d}-|1\rangle_{d})$. The initial state of the device can be rewritten as $|\phi_i\rangle_{d}=1/\sqrt{2}(|+\rangle_{d}+|-\rangle_{d})$. After the interaction, the combined device and system state is
\begin{equation}\begin{split}\label{e14}
|\Phi\rangle_{sd}=&\frac{1}{\sqrt{2}}[\alpha_1(e^{ig}|+\rangle_d|0\rangle_s+e^{-ig}|-\rangle_d|0\rangle_s) \\
&+\beta_1(e^{-ig}|+\rangle_d|1\rangle_s+e^{ig}\beta_1|-\rangle_d|1\rangle_s)].
\end{split}\end{equation}
Without postselection, after the weak interaction, the final measure device state
\begin{equation}\label{e15}
\rho_d=\frac{1}{2}\left(
\begin{array}{cc}
1 & |\alpha_1|^2e^{2ig}+|\beta_1|^2e^{-2ig} \\
|\alpha_1|^2e^{-2ig}+|\beta_1|^2e^{2ig} & 1 \\
\end{array}
\right),
\end{equation}
where the basis is $\{|+\rangle_d, |-\rangle_d\}$. The expectation value of the observable $\mathbf{O}$ of the measure device without postselection
\begin{equation}\label{e16}
\left<\mathbf{O}\right>=\text{tr}(\mathbf{O}\rho_{d})=\sin^2g.
\end{equation}
As the coupling strength is very weak $g\ll 1$, the reading of the measure device is $\left<\mathbf{O}\right>\approx g^2$ in the ordinary measurement.

In the weak measurements the postselection process is added. If the postselection state is denoted by $|\psi_f\rangle_{s}=\alpha_2|0\rangle_s+\beta_2|1\rangle_s$, the final measure device state is
\begin{equation}\label{e17}
\rho'_{d}=\frac{C_1+C_2+C_3+C_4}{2Pro},
\end{equation}
where $C_1=|\alpha_1|^2|\alpha_2|^2(|+\rangle_d\langle+|+e^{2ig}| +\rangle_d\langle-| +e^{-2ig}|-\rangle_d\langle+| +|-\rangle_d\langle-|)$, $C_2=|\beta_1|^2|\beta_2|^2(|+\rangle_d\langle+|+e^{-2ig}|+\rangle_d\langle-|+e^{2ig}|-\rangle_d\langle+|+|-\rangle_d\langle-|)$, $C_3=\alpha_1\beta^*_1\alpha^*_2\beta_2(e^{2ig}|+\rangle_d\langle+|+|+\rangle_d\langle-|+|-\rangle_d\langle+|+e^{-2ig}|-\rangle_d\langle-|)$, $C_4=\alpha^*_1\beta_1\alpha_2\beta^*_2(e^{-2ig}|+\rangle_d\langle+|+|+\rangle_d\langle-|+|-\rangle_d\langle+|+e^{2ig}|-\rangle_d\langle-|)$, and
the probability of obtaining the postselection state $|\psi_f\rangle_s$ is $Pro=|\alpha_1|^2|\alpha_2|^2+|\beta_1|^2|\beta_2|^2+2\mathbf{Re}(\alpha^*_1\beta_1\alpha_2\beta^*_2)\cos 2g$. The expectation value of $\mathbf{O}$ conditioned obtaining the postselection state $|\psi_f\rangle_s$ is
\begin{equation}\label{e18}
\left<\mathbf{O}\right>'=\frac{(|\alpha_1|^2|\alpha_2|^2+|\beta_1|^2|\beta_2|^2-2\mathbf{Re}(\alpha^*_1\beta_1\alpha_2\beta^*_2))\sin^2g} {|\alpha_1|^2|\alpha_2|^2+|\beta_1|^2|\beta_2|^2+2\mathbf{Re}(\alpha^*_1\beta_1\alpha_2\beta^*_2)\cos 2g}.
\end{equation}
The maximum readings of observable $\mathbf{O}$ in weak measurements is
\begin{equation}\label{e19}
\left<\mathbf{O}\right>'_{max}=1,
\end{equation}
which is the biggest reading of the qubit's observable $\mathbf{O}$ and the maximum readings is obtained when the preselection state $|\psi_i\rangle_{s}$ and the postselection state $|\psi_f\rangle_s$ are orthogonally. The detail derivation of Eq. (\ref{e19}) is given in Appendix B. From Eqs. (\ref{e16}) and (\ref{e19}), we can see that the strategy of weak measurement could improve the device's reading significantly by choosing appropriate PPS.

For the preselection state is the one given by Eq.(\ref{e6}) which may not be a pure state, the measure device state after the interaction is
\begin{equation}\label{e20}
\rho''_{d}=\frac{E_1+E_2+E_3+E_4}{2Pro},
\end{equation}
where $E_1=((1-r)|\alpha_2|^2/2+r|\alpha_1|^2|\alpha_2|^2)(|+\rangle_d\langle+|+e^{2ig}| +\rangle_d\langle-| +e^{-2ig}|-\rangle_d\langle+| +|-\rangle_d\langle-|)$,
$E_2=((1-r)|\beta_2|^2/2+r|\beta_1|^2|\beta_2|^2)(|+\rangle_d\langle+|+e^{-2ig}|+\rangle_d\langle-|+e^{2ig}|-\rangle_d\langle+| +|-\rangle_d\langle-|)$, $E_3=r\alpha_1\beta^*_1\alpha^*_2\beta_2(e^{2ig}|+\rangle_d\langle+|+|+\rangle_d\langle-|+|-\rangle_d\langle+|+e^{-2ig}|-\rangle_d\langle-|)$, $E_4=r\alpha^*_1\beta_1\alpha_2\beta^*_2(e^{-2ig}|+\rangle_d\langle+|+|+\rangle_d\langle-|+|-\rangle_d\langle+|+e^{2ig}|-\rangle_d\langle-|)$, and the probability of obtaining the state $|\psi_f\rangle_s$ is $Pro=(1-r)/2+r(|\alpha_1|^2|\alpha_2|^2+|\beta_1|^2|\beta_2|^2+2\mathbf{Re}(\alpha^*_1\beta_1\alpha_2\beta^*_2)\cos 2g)$. The reading of the observable is
\begin{equation}\label{e21}
\left<\mathbf{O}\right>''=\text{tr}(\mathbf{O}\rho''_{D})=\frac{F_1}{F_2},
\end{equation}
where $F_1=((1-r)/2+r(|\alpha_1|^2|\alpha_2|^2+|\beta_1|^2|\beta_2|^2-2\mathbf{Re}(\alpha^*_1\beta_1\alpha_2\beta^*_2)))\sin^2g$, and $F_2=(1-r)/2+r(|\alpha_1|^2|\alpha_2|^2+|\beta_1|^2|\beta_2|^2+2\mathbf{Re}(\alpha^*_1\beta_1\alpha_2\beta^*_2)\cos2g)$. By choosing orthogonal PPS, we get the maximum reading
\begin{equation}\label{e22}
\left<\mathbf{O}\right>''_{max}=\frac{(1+r)\sin^2g}{(1-r)+2r\sin^2g}.
\end{equation}
The detail derivation of Eq. (\ref{e22}) is given in Appendix C, and the conditions of obtaining maximum shifts are the PPS orthogonal. The visual relationships between the modulus $r$ of the postselection state and the maximum reading of the measure device are showed in Fig. 2.
\begin{figure}[t]
\centering \includegraphics[scale=0.5]{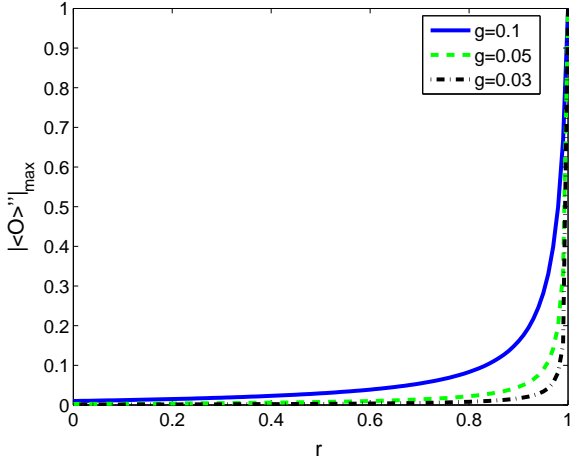} \caption{(Color online) For the qubit measure device, the relationships between the maximum readings
and the modulus $r$ with $g=0.1$, $g=0.05$, and$g=0.03$.}
\label{fig:2}
\end{figure}

From Fig. 2, the maximum readings of the measure device decrease with the modulus $r$ very sharply. When the $r=0$, the maximum reading is the same with the one given by Eq. (\ref{e16}) obtained in ordinary measurement. In this section, we have the statement that if we want obtain significant amplification effects in weak measurements, the preselection state should be kept in pure state regardless of the types of the measure devices.

\section{The quantum noise and the amplification}
In this section, we study the relationships between the strength of the quantum noise and the maximum readings of the measure device. Three typical quantum noises: depolarizing channel, phase damping, and amplitude damping are considered.
\subsection{The depolarizing channel and the the amplification}
\begin{equation}\begin{split}\label{e24}
|\delta p'|_{max}=\frac{g}{\sqrt{1-(1-\gamma)^2e^{-4\Delta^2g^2}}},\\
|\delta q'|_{max}=\frac{2(1-\gamma)g\Delta^2e^{-2\Delta^2g^2}}{\sqrt{1-(1-\gamma)^2e^{-4\Delta^2g^2}}}.
\end{split}\end{equation}
\begin{figure}[t]
\centering \includegraphics[scale=0.5]{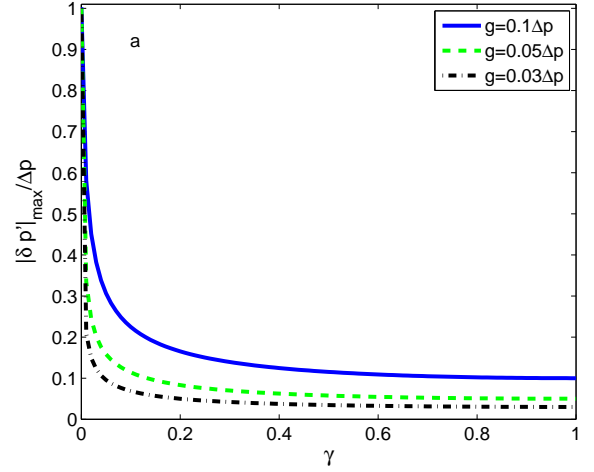} \includegraphics[scale=0.5]{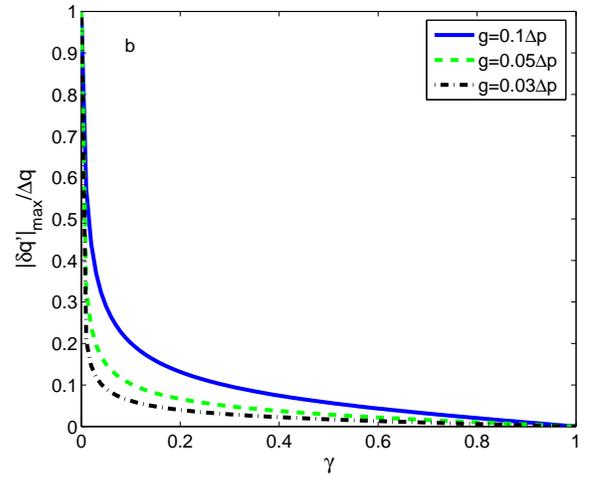} \caption{(Color online) The relationship between the maximum shifts
and the noise strength $\gamma$ of depolarizing channel with $g=0.1\Delta p$, $g=0.05\Delta p$, and$g=0.03\Delta p$.}
\label{fig:3}
\end{figure}
When the quantum systems suffer from a depolarizing channel~\cite{nielsen}, the pure preselection state $|\psi_i\rangle_s$ will evolve into
\begin{equation}\label{e23}
\rho'_s=\frac{\gamma \mathbf{I}}{2}+(1-\gamma)|\psi_i\rangle_{s}\langle\psi_i|,
\end{equation}
where $\gamma$ is the probability that quantum system is depolarized, and can also be regarded as the strength of the quantum noise. The modulus of the state $\rho'_s$ is $1-\gamma$, if the measure device is Gaussain type given by Eq. (\ref{e7}), and the interaction is described by Eq. (\ref{e8}), from Eq. (\ref{e11}), we get the maximum shifts under the depolarizing channel

The relationships between the maximum shifts and the noise strength $\gamma$ are pictured in Fig. 3.
In Fig. 3, it is shown that the maximum readings decrease quickly with the increase of the noise strength $\gamma$. If we want to obtain significant amplification effects, the preselection quantum systems must be kept away from the depolarizing channel. For example, if we want to obtain the twice maximum shifts of the ordinary measurement, $|\delta p'|_{max}\geq 2g$, the noise strength $\gamma\leq 0.13$.

For the measure device is the qubit state described by Eq. (\ref{e12}), and the interaction between the measure device and the systems is given by Eq. (\ref{e13}), as the modulus of state $\rho'_s$ in Eq. (\ref{e23}) is $1-\gamma$, using the result obtained in Eq. (\ref{e22}), we have the maximum reading of the qubit measure device under the depolarizing channel
\begin{equation}\label{e25}
\left<\mathbf{O}'\right>_{max}=\frac{(2-\gamma)\sin^2g}{\gamma+2(1-\gamma)\sin^2g}.
\end{equation}
The relationships between the maximum readings of qubit measure device and the noise strength are pictured in Fig. 4.
\begin{figure}[t]
\centering \includegraphics[scale=0.5]{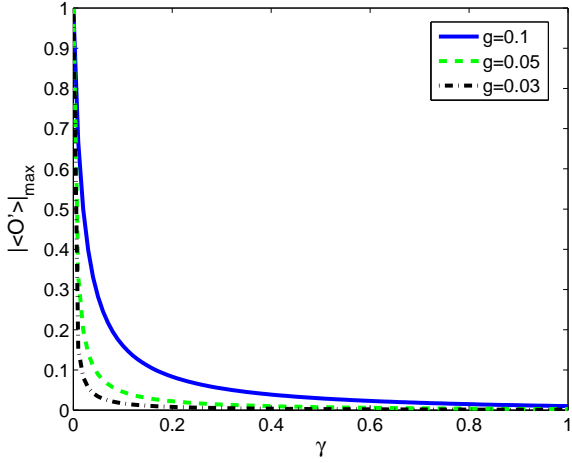} \caption{(Color online) The relationships between the maximum readings
and the noise strength $\gamma$ of depolarizing channel with $g=0.1\Delta p$, $g=0.05\Delta p$, and$g=0.03\Delta p$.}
\label{fig:4}
\end{figure}
Fig. 4 gives the similar results of Fig. 3. Then we have a statement that for obtaining significant amplification effects of weak measurement, the strength of the depolarizing channel noise should be very weak regardless of the types of measure devices.

\subsection{The phase damping and the the amplification}
If a quantum state $\rho$ is disturbed by the phase damping noise, the final quantum state will evolves into~\cite{nielsen}
\begin{equation}\label{e26}
\rho'=\mathbf{E}_0\rho \mathbf{E}^{\dagger}_0+\mathbf{E}_1\rho \mathbf{E}^{\dagger}_1,
\end{equation}
where
\begin{equation}\label{e27}
\mathbf{E}_0=\left(
\begin{array}{cc}
1 & 0 \\
0 & 1-\gamma\\
\end{array}
\right),
\mathbf{E}_1=\left(
\begin{array}{cc}
0 & 0 \\
0 & \gamma\\
\end{array}
\right),
\end{equation}
and the $\gamma$ can be seen as the strength of the phase damping noise. The initial preselection state is $|\psi_i\rangle_s=\alpha_1|0\rangle_s+\beta_1|1\rangle_s$, under the phase damping, $|\psi_i\rangle_s$ evolves into
\begin{equation}\label{e28}
\rho'_{s}=\left(
\begin{array}{cc}
|\alpha_1|^2&(1-\gamma)\alpha_1\beta^*_1\\
(1-\gamma)\alpha^*_1\beta_1 & |\beta_1|^2\\
\end{array}
\right)
\end{equation}
with the basis $\{|0\rangle_s,|1\rangle_s\}$.

 When the measure device state is Gaussian type given by Eq. (\ref{e7}), and the interaction Hamiltonian is described in Eq. (\ref{e8}), conditioned on obtaining postselection state $|\psi_f\rangle_s$, the final measure device state is
\begin{equation}\label{e29}
\rho'_{d}=\frac{G_1+G_2+G_3+G_4}{Pro}
\end{equation}
where $G_1 = |\alpha_1|^2 |\alpha_2|^2e^{igq} \rho_{d}e^{-igq}$, $G_2 = |\beta_1|^2|\beta_2|^2 e^{-igq} \rho_{d}e^{igq}$,
$G_3=(1-\gamma)\alpha_1^*\beta_1\alpha_2\beta_2^*e^{-igq}\rho_{d}e^{-igq}$, $G_4=(1-\gamma)\alpha_1\beta_1^*\alpha_2^*\beta_2e^{igq}\rho_{d}e^{igq}$, $\rho_{d}={1}/{\sqrt{2\pi\Delta^2}}\iint e^{-q_1^2/{4\Delta^2}}e^{-q_2^2/{4\Delta^2}}|q_1\rangle\langle q_2|\text{d}q_1\text{d}q_2$ given by Eq. (\ref{e7}), and $Pro=|\alpha_1|^2|\alpha_2|^2+|\beta_1|^2|\beta_2|^2+2(1-\gamma)\mathbf{Re}(\alpha_1^*\beta_1\alpha_2\beta_2^*)e^{-2\Delta^2g^2}$ is the probability of obtaining the postselection state. The shifts of the pointer momentum and position are
\begin{equation}\begin{split}\label{e30}
\delta p'&=\frac{(|\alpha_1|^2|\alpha_2|^2-|\beta_1|^2|\beta_2|^2)g}{Pro},\\
\delta q'&=\frac{4(1-\gamma)g\Delta^2\mathbf{Im}(\alpha_1^*\beta_1\alpha_2\beta_2^*)}{Pro}.
\end{split}\end{equation}
From the derivation given in Appendix D, we obtain the maximum shifts by choosing appropriate PPS
\begin{equation}\begin{split}\label{e31}
|\delta p'|_{max}&=\frac{g}{\sqrt{1-(1-\gamma)^2e^{-4\Delta^2g^2}}},\\
|\delta q'|_{max}&=\frac{2(1-\gamma)g\Delta^2e^{-2\Delta^2g^2}}{\sqrt{1-(1-\gamma)^2e^{-4\Delta^2g^2}}}.
\end{split}\end{equation}
And the conditions of attaining maximum shifts are also given in Appendix D. Comparing this equation with Eq. (\ref{e24}), we can see that the relationships between maximum shifts and noise strength are completely same for the depolarizing channel and phase damping noises. And the visual relationships are also shown in Fig. 3.

If the measure device is the qubit state described in Eq. (\ref{e12}), and the interaction is given in Eq.(\ref{e13}), conditioned on obtaining $|\psi_f\rangle_s$, the final device state is
\begin{equation}\label{e32}
\rho'_{D}=\frac{J_1+J_2+J_3+J_4}{2Pro},
\end{equation}
where $J_1=|\alpha_1|^2|\alpha_2|^2(|+\rangle_d\langle+|+e^{2ig}| +\rangle_d\langle-| +e^{-2ig}|-\rangle_d\langle+| +|-\rangle_d\langle-|)$, $J_2=|\beta_1|^2|\beta_2|^2(|+\rangle_d\langle+|+e^{-2ig}|+\rangle_d\langle-|+e^{2ig}|-\rangle_d\langle+|+|-\rangle_d\langle-|)$, $J_3=(1-\gamma)\alpha_1\beta^*_1\alpha^*_2\beta_2(e^{2ig}|+\rangle_d\langle+|+|+\rangle_d\langle-|+|-\rangle_d\langle+|+e^{-2ig}|-\rangle_d\langle-|)$, $J_4=(1-\gamma)\alpha^*_1\beta_1\alpha_2\beta^*_2(e^{-2ig}|+\rangle_d\langle+|+|+\rangle_d\langle-|+|-\rangle_d\langle+|+e^{2ig}|-\rangle_d\langle-|)$, and
the probability of obtaining the poseselection state is $Pro=|\alpha_1|^2|\alpha_2|^2+|\beta_1|^2|\beta_2|^2+2(1-\gamma)\mathbf{Re}(\alpha^*_1\beta_1\alpha_2\beta^*_2)\cos 2g$. The reading of the observable $\mathbf{O}$ is
\begin{equation}\label{e33}
\left<\mathbf{O}'\right>=\frac{N_1}{N_2}.
\end{equation}
where $N_1=(|\alpha_1|^2|\beta_2|^2+|\beta_1|^2|\beta_2|^2-2(1-\gamma)\mathbf{Re}(\alpha_1^*\beta_1\alpha_2\beta_2^*))\sin^2g$ and $N_2=|\alpha_1|^2|\alpha_2|^2+|\beta_1|^2|\beta_2|^2+2(1-\gamma)\mathbf{Re}(\alpha^*_1\beta_1\alpha_2\beta^*_2)\cos 2g$.
By choosing orthogonal PPS, the maximum readings of qubit measure device is
\begin{equation}\label{e34}
\left<\mathbf{O}'\right>_{max}=\frac{(2-\gamma)\sin^2g}{\gamma+2(1-\gamma)\sin^2g}.
\end{equation}
The derivation of this equation is given in Appendix E. This maximum reading is the same with the one given in Eq. (\ref{e25}), and is also visualized in Fig . 4. From Figs. 3 and 4, the maximum readings decrease sharply with the increase of the strength of the depolarizing channel and phasing damping noises. If we want to obtain remarkable amplified readings to improve measure precision, the quantum systems must be away from the depolarizing channel and phase damping noises.
\subsection{The amplitude damping and the amplification}
If a quantum systems $\rho$ undergoes a amplitude damping operation from outside environment, the state will evolve into~\cite{nielsen}
\begin{equation}\label{e35}
\rho'=\mathbf{E}_0 \rho \mathbf{E}_0^{\dagger}+\mathbf{E}_1\rho\mathbf{E}_1^{\dagger},
\end{equation}
where
\begin{equation}\label{e36}
\mathbf{E}_0=\left(
\begin{array}{cc}
1 & 0 \\
0 & \sqrt{1-\gamma}\\
\end{array}
\right),
\mathbf{E}_1=\left(
\begin{array}{cc}
0 & \sqrt{\gamma} \\
0 & 0\\
\end{array}
\right),
\end{equation}
and $\gamma$ can be regarded as the noise strength of the amplitude damping. Under the amplitude damping, the preselection state $|\psi_i\rangle_s=\alpha_1|0\rangle_s+\beta_1|1\rangle_s $ evolves into
\begin{equation}\label{e37}
\rho'_{s}=\left(
\begin{array}{cc}
|\alpha_1|^2+\gamma|\beta_1|^2&\sqrt{1-\gamma}\alpha_1\beta^*_1\\
\sqrt{1-\gamma}\alpha^*_1\beta_1 &(1-\gamma) |\beta_1|^2\\
\end{array}
\right),
\end{equation}
and the basis is $\{|0\rangle_s,|1\rangle_s\}$.

\begin{figure}[t]
\centering \includegraphics[scale=0.22]{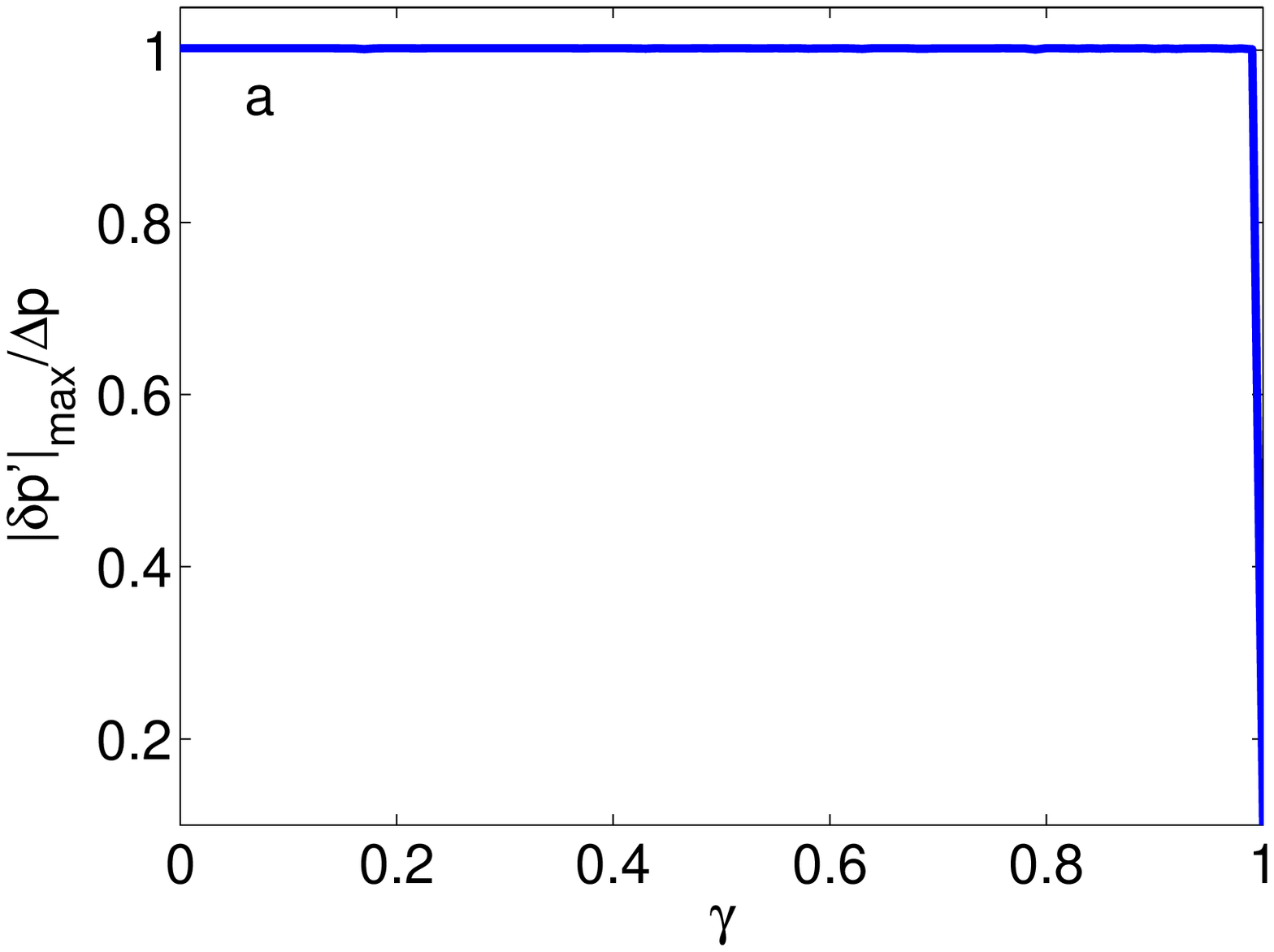} \includegraphics[scale=0.22]{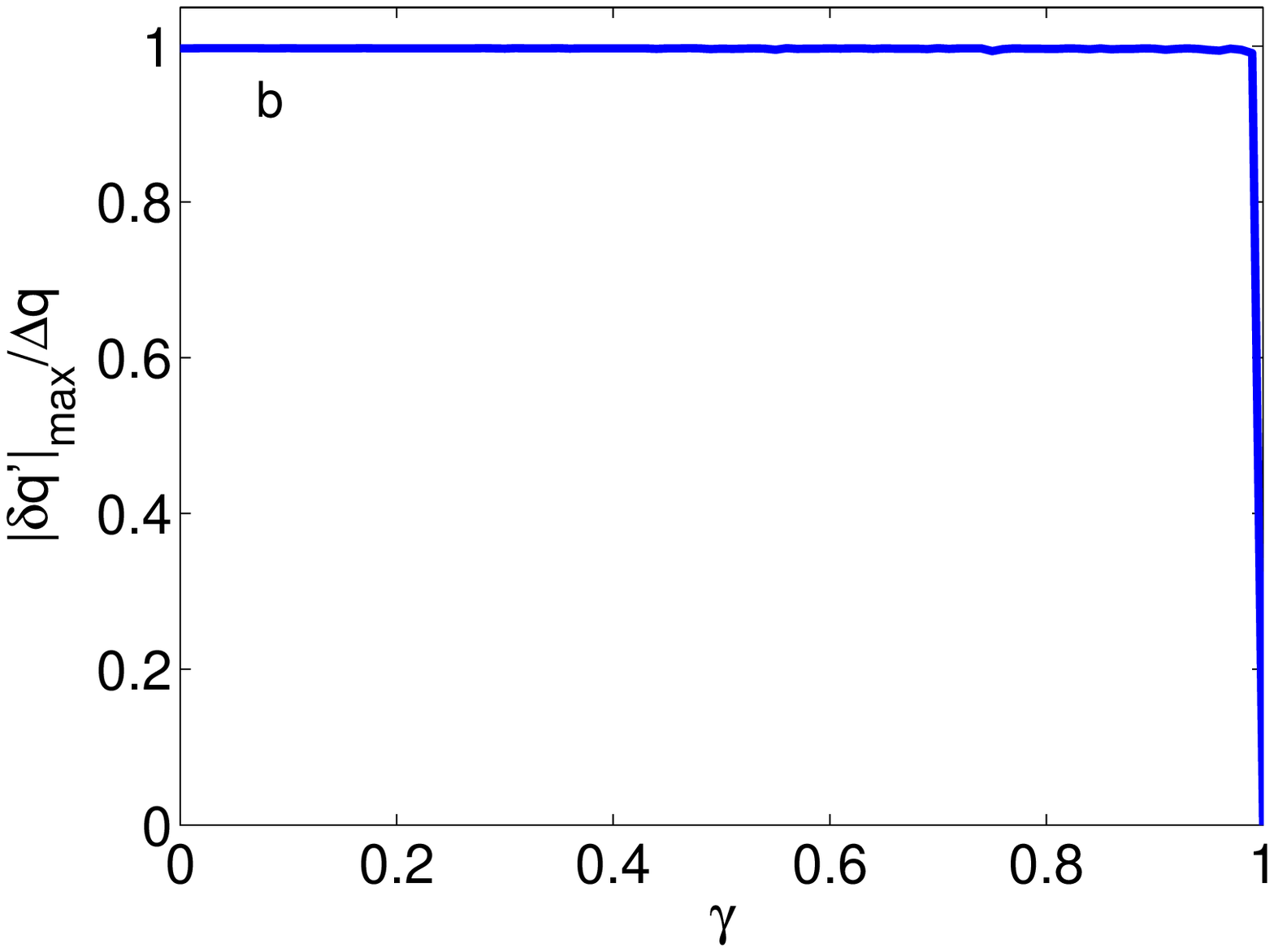} \caption{(Color online) The relationship between the maximum shifts
and the noise strength $\gamma$ of depolarizing channel with $g=0.1\Delta p$.}
\label{fig:5}
\end{figure}

When the measure device state is the Gaussian type given by Eq. (\ref{e7}), and the interaction is described in Eq. (\ref{e8}), the final measure device state is
\begin{equation}\label{e38}
\rho'_d=\frac{K_1+K_2+K_3+K_4}{Pro},
\end{equation}
where $K_1 = (|\alpha_1|^2 +|\beta_1|^2 \gamma)|\alpha_2|^2 e^{igq} \rho_{d}e^{-igq}$, $K_2 = (1-\gamma)|\beta_1|^2|\beta_2|^2 e^{-igq} \rho_{d}e^{igq}$,
$K_3 = \sqrt{1-\gamma}\alpha_1^*\beta_1\alpha_2\beta_2^*e^{-igq}\rho_{d}e^{-igq}$, $K_4 = \sqrt{1-\gamma}\alpha_1\beta_1^*\alpha_2^*\beta_2 e^{igq}\rho_{d}e^{igq}$, $\rho_{d} = {1}/{\sqrt{2\pi\Delta^2}}\iint e^{-q_1^2/{4\Delta^2}}e^{-q_2^2/{4\Delta^2}}|q_1\rangle\langle q_2|\text{d}q_1\text{d}q_2$ is the initial measure device state, and $Pro=(|\alpha_1|^2+|\beta_1|^2\gamma)|\alpha_2|^2+(1-\gamma)|\beta_1|^2|\beta_2|^2+2\sqrt{1-\gamma}\mathbf{Re}(\alpha_1^*\beta_1\alpha_2\beta_2^*)e^{-2\Delta^2g^2}$ is the probability of obtaining the postselection state $|\psi_f\rangle_s$. The shifts of the pointer are
\begin{equation}\begin{split}\label{e39}
\delta p '&=\frac{|\alpha_2|^2(|\alpha_1|^2 +\gamma|\beta_1|^2)-(1-\gamma)|\beta_2|^2|\beta_1|^2}{Pro}g,\\
\delta q' &=\frac{4\sqrt{1-\gamma}g\Delta^2e^{-2g^2\Delta^2}\mathbf{Im}(\alpha_1^*\beta_1\alpha_2\beta_2^*)}{Pro}.
\end{split}\end{equation}

However, we have not gotten the analytical expressions of the maximum shifts under the amplitude damping, we obtain the numerical maximum shifts in Fig 5. From Fig. 5, the maximum shifts are the same with the ones obtained in the weak measurements without quantum noises by selecting appropriate PPS except that the noise strength is 1. Under the amplitude phasing noise, the significant amplification of weak measurements could also be obtained.

When the measure device is a qubit whose state is given by Eq. (\ref{e12}), and the interaction between the device and systems is given by Eq. (\ref{e13}), the final measure device state is
\begin{equation}\label{e40}
\rho'_{d}=\frac{L_1+L_2+L_3+L_4}{Pro},
\end{equation}
where $L_1 = (|\alpha_1|^2 +|\beta_1|^2 \gamma)|\alpha_2|^2 e^{ig\sigma_x} \rho_{d}e^{-ig\sigma_x}$, $L_2 = (1-\gamma)|\beta_1|^2|\beta_2|^2 e^{-ig\sigma_x} \rho_{d}e^{ig\sigma_x}$,
$L_3 = \sqrt{1-\gamma}\alpha_1^*\beta_1\alpha_2\beta_2^*e^{-ig\sigma_x}\rho_{d}e^{-ig\sigma_x}$, $L_4 = \sqrt{1-\gamma}\alpha_1\beta_1^*\alpha_2^*\beta_2 e^{ig\sigma_x}\rho_{d}e^{ig\sigma_x}$, $\rho_{d} = {1}/2(|+\rangle_d\langle+|+| +\rangle_d\langle-| +|-\rangle_d\langle+| +|-\rangle_d\langle-|)$ is the initial measure device state, and $Pro=(|\alpha_1|^2+|\beta_1|^2\gamma)|\alpha_2|^2+(1-\gamma)|\beta_1|^2|\beta_2|^2+2\sqrt{1-\gamma}\mathbf{Re}(\alpha_1^*\beta_1\alpha_2\beta_2^*)\cos2g$ is the probability of obtaining the postselection state $|\psi_f\rangle_s$.
\begin{figure}[t]
\centering \includegraphics[scale=0.45]{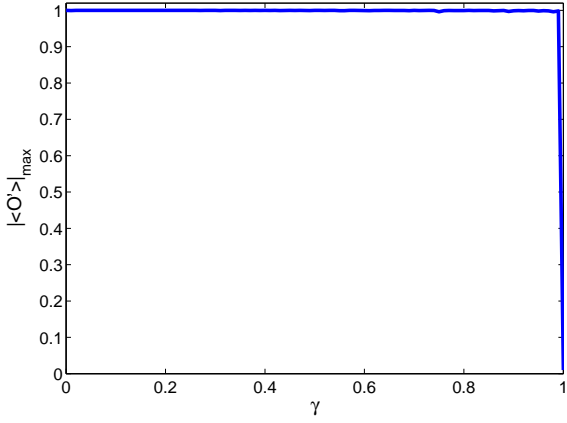} \caption{(Color online) For the qubit measure device, the relationship between the maximum readings
and the coupling strength $\gamma$ of the amplitude phasing with $g=0.1$.}
\label{fig:6}
\end{figure}
The reading of the measure device conditioned on obtaining the poselection state is
\begin{equation}\label{e41}
\left<\mathbf{O}'\right>=\frac{M_1}{M_2}\sin^2g,
\end{equation}
where $M_1=(|\alpha_1|^2+\gamma|\beta_1|^2)|\alpha_2|^2+(1-\gamma)|\beta_1|^2|\beta_2|^2-2\sqrt{1-\gamma}\mathbf{Re}(\alpha_1^*\beta_1\alpha_2\beta_2^*)$, $M_2=|\alpha_1|^2|\alpha_2|^2+\gamma|\beta_1|^2|\alpha_2|^2+(1-\gamma)|\beta_1|^2|\beta_2|^2+2\sqrt{1-\gamma}\mathbf{Re}(\alpha_1^*\beta_1\alpha_2\beta_2^*)\cos2g$.
We have not obtained the analytical maximum reading of observable $\mathbf{O}$, the numerical maximum reading are pictured in Fig. 6. From Fig. 6, the maximum value $|\left<\mathbf{O}'\right>|_{max}$ is the same with the one without quantum noise for $\gamma \neq 1$. To obtain significantly amplified shifts by using weak measurement, we should not worry about the amplitude phasing when the noise strength $\gamma\neq 1$.

\section{conclusion}
For Gaussian and qubit measure devices, the maximum shifts of measure device are discussed in this article. We have the statements that the maximum shifts decrease quickly with the modulus of the preselection states, and the maximum readings decrease sharply with strength of the depolarizing channel and phase damping noises. If we want to attain significant amplified shifts in weak measurements, the preselection quantum systems must be ensured without the disturbances of the two typical noises. For the amplitude damping, it is shown that the maximum shifts are the same with the ones without noise if the coupling strength is not equal to 1. We can also obtain significant amplification under the amplitude damping by choosing orthogonal PPS. The conditions of obtaining maximum shifts are also given in this article. Those results may be helpful for experimentalists to implement high precise measurements.

\section*{Acknowledgments}
This work was financially supported by the National Natural Science Foundation of China (Grants No. 11305118, and No. 11175063), and the Fundamental Research Funds
for the Central Universities.
\section{Appendix}
\subsection{The derivation of Equation (\ref{e11})}

The pure two dimensional states $|\psi_i\rangle_s$ and $|\psi_f\rangle_s$ can be rewritten as $|\psi_i\rangle_s=\cos\frac{\theta_1}{2}|0\rangle_s+e^{i\phi_1}\sin\frac{\theta_1}{2}|1\rangle_s$ and $|\psi_f\rangle_s = \cos\frac{\theta_2}{2}|0\rangle_s+e^{i\phi_2}\sin\frac{\theta_2}{2}|1\rangle_s$. The shifts of momentum and position in Eq. (\ref{e10}) could be rewritten
\begin{equation}\begin{split}\label{e42}
\delta p'=\frac{g(\cos\theta_2+r\cos\theta_1)}{1+r\cos\theta_1\cos\theta_2+r\sin\theta_1\sin\theta_2\cos\phi_0e^{-2\Delta^2g^2}},\\
\delta q'=\frac{2gr\Delta^2\sin\theta_1\sin\theta_2\sin\phi_0}{1+r\cos\theta_1\cos\theta_2+r\sin\theta_1\sin\theta_2\cos\phi_0e^{-2\Delta^2g^2}},
\end{split}\end{equation}
where $\phi_0=\phi_1-\phi_2$.

First, we search for the extreme values of $\delta p'$. For $\frac{\partial \delta p' }{\partial \phi_0}=0$, we get $\sin \phi_0=0$, $\cos\phi_0=\pm 1$ and
\begin{equation}\label{e43}
\delta p'=\frac{g(\cos\theta_2+r\cos\theta_1)}{1+r\cos\theta_1\cos\theta_2\pm r\sin\theta_1\sin\theta_2e^{-2\Delta^2g^2}}\\
\end{equation}
Next, the method of Lagrange multipliers is used. Let $x_1=\cos\theta_1$, $x_2=\cos\theta_1$, $y_1=\sin\theta_1$, and $y_2=\sin\theta_2$ with the constraint
$x_1^2+y_1^2-1=0$ and $x_2^2+y_2^2-1=0$, we have the Lagrange function
\begin{equation}\begin{split}\label{e44}
F_1=&\frac{(x_1+rx_2)g}{1+rx_1x_2\pm r y_1y_2e^{-2\Delta^2g^2}}\\
&+\lambda_1(x_1^2+y_1^2-1)+\lambda_2(x_2^2+y_2^2-1).
\end{split}\end{equation}
Let $\frac{\partial F_1}{\partial x_1}=\frac{\partial F_1}{\partial y_1}=0$, from $y_1\frac{\partial F_1}{\partial x_1}=x_1\frac{\partial F_1}{\partial y_1}$, we have $y_1=0$ or
\begin{equation}\label{e45}
y_1y_2=\mp(r+x_1x_2)e^{-2\Delta^2g^2}.
\end{equation}
For $y_1=0$, we have the shift
\begin{equation}\label{e46}
 \delta p '=\pm g.
 \end{equation}
Let $\frac{\partial F_1}{\partial x_2}=\frac{\partial F_1}{\partial y_2}=0$, from $y_2\frac{\partial F_1}{\partial x_2}=x_2\frac{\partial F_1}{\partial y_2}$, we get
\begin{equation}\label{e47}
\frac{y_1}{y_2}=\mp \frac{r e^{-2\Delta^2g^2}(1+rx_1x_2)}{1-r^2x_2^2}.
\end{equation}
From Eqs. (\ref{e45}) and (\ref{e47}), we have
\begin{equation}\label{e48}
y_2^2=\frac{(r+x_1x_2)(1-r^2x_2^2)}{r(1+rx_1x_2)}.
\end{equation}
Using the constraint $x_2^2+y_2^2=1$, we have
\begin{equation}\label{e49}
x_2=-\frac{x_1}{r}~or~x_2=0.
\end{equation}
For $x_2=-\frac{x_1}{r}$, we have
\begin{equation}\label{e50}
\delta p'= 0.
\end{equation}
For $x_2=0$, we get $y_2=\pm 1$, and from Eq. (\ref{e45}), we have $y_1=\mp e^{-2\Delta^2g^2}$, $x_1=\pm\sqrt{1-r^2e^{-4\Delta^2g^2}}$, and
\begin{equation}\label{e51}
\delta p'=\pm\frac{g}{\sqrt{1-r^2e^{-4\Delta^2g^2}}}.
\end{equation}
From Eqs. (\ref{e46}), (\ref{e50}), and (\ref{e51}), we have the maximum shift of momentum
\begin{equation}\label{e52}
|\delta p'|_{max}=\frac{g}{\sqrt{1-r^2e^{-4\Delta^2g^2}}}.
\end{equation}
The conditions of obtaining the maximum shift is $\cos\theta_2=0$, $\sin\theta_2=\pm 1$, $\cos\theta_1=\pm re^{-2\Delta^2g^2}$, and $\sin\theta_1=\pm\sqrt{1-r^2e^{-4\Delta^2g^2}}$.

Now we search for the maximum value of $|\delta q'|$ using the method of Lagrange multipliers. Let $x_3=\cos\phi_0$, and $y_3=\sin\phi_0$, we have the the Lagrange function
\begin{equation}\begin{split}\label{e53}
F_2=&\frac{2gr\Delta^2y_1y_2y_3}{1+rx_1x_2+ry_1y_2x_3e^{-2\Delta^2g^2}}+\lambda_1(x_1^2+y_1^2-1)\\
&+\lambda_2(x_2^2+y_2^2-1)+\lambda_3(x_3^2+y_3^2-1).
\end{split}\end{equation}
For $\frac{\partial F_2}{\partial x_1}=\frac{\partial F_2}{\partial y_1}=0$, from $y_1\frac{\partial F_2}{\partial x_1}=x_1\frac{\partial F_2}{\partial y_1}$, we have
\begin{equation}\label{e54}
y_2y_3(x_1+rx_2)=0.
\end{equation}
The solution of the equation is $y_2=0$, or $y_3=0$, or
\begin{equation}\label{e55}
x_1=-rx_2.
\end{equation}
For $y_2=0$ or $y_3=0$, we have
\begin{equation}\label{e56}
\delta q'=0.
\end{equation}
For $\frac{\partial F_2}{\partial x_2}=\frac{\partial F_2}{\partial y_2}=0$, from $y_2\frac{\partial F_2}{\partial x_2}=x_2\frac{\partial F_2}{\partial y_2}$, we get $y_2=0$, or $y_3=0$, or
\begin{equation}\label{e57}
 x_1=-rx_2.
\end{equation}
From Eqs (\ref{e55}) and (\ref{e57}), as $r$ is not always equal to 1, we have $x_1=x_2=0$ and $y_1y_2=\pm 1$. And the Lagrange function changes into
\begin{equation}\label{e58}
F_2(x_3,y_3,\lambda_3)=\pm \frac{2gr\Delta^2y_3}{1\pm rx_3e^{-2\Delta^2g^2}}.
\end{equation}
For $\frac{\partial F_2}{\partial x_3}=\frac{\partial F_2}{\partial y_3}=0$, from $y_3\frac{\partial F_2}{\partial x_3}=x_3\frac{\partial F_2}{\partial y_3}$, we have
\begin{equation}\label{e59}
x_3=\mp re^{-2\Delta^2g^2}, y_3=\pm\sqrt{1-r^2e^{-4\Delta^2g^2}}.
\end{equation}
And the shifts of position is
\begin{equation}\label{e60}
\delta q'=\pm\frac{2gr\Delta^2}{\sqrt{1-r^2e^{-4\Delta^2g^2}}}.
\end{equation}
From Eqs. (\ref{e56}) and (\ref{e60}), we get the maximum position shift
\begin{equation}\label{e61}
|\delta q'|_{max}=\frac{2gr\Delta^2}{\sqrt{1-r^2e^{-4\Delta^2g^2}}}.
\end{equation}
The conditions of attaining the maximum shift is $\sin\theta_1=\pm 1$, $\sin\theta_1=\mp 1$, $\cos\theta_3=\mp re^{-2\Delta^2g^2}$.
\subsection{The derivation of Equation (\ref{e19})}
For the two pure states $|\psi_i\rangle_s=\cos\frac{\theta_1}{2}|0\rangle_s+e^{i\phi_1}\sin\frac{\theta_1}{2}|1\rangle_s$ and $|\psi_f\rangle_s = \cos\frac{\theta_2}{2}|0\rangle_s+e^{i\phi_2}\sin\frac{\theta_2}{2}|1\rangle_s$, the Eq. (\ref{e18}) could be rewritten as
\begin{equation}\label{e62}
\left<\mathbf{O}'\right>=\frac{(1+\cos\theta_1\cos\theta_2-\sin\theta_1\sin\theta_2\cos\phi_0)\sin^2g}{1+\cos\theta_1\cos\theta_2+\sin\theta_1\sin\theta_2\cos\phi_0\cos2g},
\end{equation}
where $\phi_0=\phi_1-\phi_2$. From $\frac{\partial \left<\mathbf{O}'\right>}{\partial \phi_0}=0$, we have $\sin\phi_0=0$, $\cos\phi_0=\pm 1$, and
\begin{equation}\label{e63}
\left<\mathbf{O}'\right>=\frac{(1+\cos\theta_1\cos\theta_2\mp\sin\theta_1\sin\theta_2)\sin^2g}{1+\cos\theta_1\cos\theta_2\pm\sin\theta_1\sin\theta_2\cos2g}.
\end{equation}
Let $t=\cos\frac{\theta_1\mp\theta_2}{2}/\cos\frac{\theta_1\pm\theta_2}{2}$, we have
\begin{equation}\label{e64}
\left<\mathbf{O}'\right>=\frac{\sin^2g}{1+t^2+(t^2-1)\cos2g}.
\end{equation}
From $\frac{\partial \left<\mathbf{O}'\right>}{\partial t}=0$, we have $t=0$, and the maximum reading
\begin{equation}\label{e65}
\left<\mathbf{O}'\right>_{max}=\frac{\sin^2g}{1-\cos2g}=1.
\end{equation}
The conditions of obtaining the maximum reading are $\phi_1-\phi_2=0$ and $\theta_1-\theta_2=\pi$, or $\phi_1-\phi_2=\pi$, and $\theta_1+\theta_2=\pi$. In other words, when the preselection and postselection states are orthogonal, the reading of qubit measure device attains its maximum value.
\subsection{The derivation of Equation (\ref{e22})}
For $|\psi_i\rangle_s=\cos\frac{\theta_1}{2}|0\rangle_s+e^{i\phi_1}\sin\frac{\theta_1}{2}|1\rangle_s$ and $|\psi_f\rangle_s= \cos\frac{\theta_2}{2}|0\rangle_s+e^{i\phi_2}\sin\frac{\theta_2}{2}|1\rangle_s$, the Eq. (\ref{e21}) could be rewritten as
\begin{equation}\label{e66}
\left<\mathbf{O}''\right>=\frac{(1-r+r(1+\cos\theta_1\cos\theta_2-\sin\theta_1\sin\theta_2\cos\phi_0))\sin^2g} {1-r+r(1+\cos\theta_1\cos\theta_2+\sin\theta_1\sin\theta_2\cos\phi_0\cos2g)},
\end{equation}
where $\phi_0=\phi_1-\phi_2$. From $\frac{\partial \left<\mathbf{O}''\right>}{\partial \phi_0}=0$, we have $\sin\phi_0=0$, $\cos\phi_0=\pm 1$, and
\begin{equation}\label{e67}
\left<\mathbf{O}''\right>=\frac{(1-r+2r\cos^2\frac{\theta_1\pm\theta_2}{2})\sin^2g}{1-r+2r(\cos^2\frac{\theta_1\mp\theta_2}{2}-\sin^2g(\cos^2\frac{\theta_1\pm\theta_2}{2}-\cos^2\frac{\theta_1\mp\theta_2}{2}))}.
\end{equation}
Since the $\left<\mathbf{O}''\right>\geq 0$ is the monotonically decreasing function with $\cos^2\frac{\theta_1\mp\theta_2}{2}$. Let $\cos^2\frac{\theta_1\mp\theta_2}{2}=0$, we obtain the maximum reading of qubit measure device
\begin{equation}\label{e68}
\left<\mathbf{O}''\right>_{max}=1.
\end{equation}
The maximum reading is obtained under the conditions $\phi_1-\phi_2=0$ and $\theta_1-\theta_2=\pi$, or $\phi_1-\phi_2=\pi$, and $\theta_1+\theta_2=\pi$ for which cases the PPS are orthogonal.
\subsection{The derivation of Equation (\ref{e31})}
For $|\psi_i\rangle_s=\cos\frac{\theta_1}{2}|0\rangle_s+e^{i\phi_1}\sin\frac{\theta_1}{2}|1\rangle_s$ and $|\psi_f\rangle_s= \cos\frac{\theta_2}{2}|0\rangle_s+e^{i\phi_2}\sin\frac{\theta_2}{2}|1\rangle_s$, the Eq. (\ref{e30}) could be rewritten as
\begin{equation}\begin{split}\label{e69}
\delta p'=\frac{(\cos\theta_1+\cos\theta_2)g}{1+\cos\theta_1\cos\theta_2+(1-\gamma)\sin\theta_1\sin\theta_2\cos\phi_0e^{-2\Delta^2g^2}},\\
\delta q'= \frac{2(1-\gamma)g\Delta^2e^{-2\Delta^2g^2}\sin\theta_1\sin\theta_2\sin\phi_0} {1+\cos\theta_1\cos\theta_2+(1-\gamma)\sin\theta_1\sin\theta_2\cos\phi_0e^{-2\Delta^2g^2}}
\end{split}\end{equation}
where $\phi_0=\phi_1-\phi_2$.
First we search for the extreme values of $\delta p'$, by $\frac{\partial \delta p' }{\partial \phi_0}=0$, we get $\sin \phi_0=0$, $\cos\phi_0=\pm 1$ and
\begin{equation}\label{e70}
\delta p'=\frac{2gt} {1+t^2\pm(1-\gamma)(1-t^2)e^{-2\Delta^2g^2}},
\end{equation}
where $t={\cos\frac{\theta_1-\theta_2}{2}}/{\cos\frac{\theta_1+\theta_2}{2}}$. From $\frac{\partial \delta p'}{\partial t}=0$, we have
\begin{equation}\label{e71}
t^2=\frac{1\pm(1-\gamma)e^{-2\Delta^2g^2}}{1\mp (1-\gamma)e^{-2\Delta^2g^2}}.
\end{equation}
Then we get the maximum shift of momentum is
\begin{equation}\label{e72}
|\delta p'|_{max}=\frac{g}{\sqrt{1-(1-\gamma)e^{-4\Delta^2g^2}}},
\end{equation}
and the conditions of obtaining the extreme values are $\phi_1-\phi_2=0$ and ${\cos\frac{\theta_1+\theta_2}{2}}/{\cos\frac{\theta_1+\theta_2}{2}}=\pm\sqrt{1+(1-\gamma)e^{-4\Delta^2g^2}/{1-(1+\gamma)e^{-4\Delta^2g^2}}}$, or
$\phi_1-\phi_2=\pi$ and ${\cos\frac{\theta_1+\theta_2}{2}}/{\cos\frac{\theta_1+\theta_2}{2}}=\pm\sqrt{1-(1-\gamma)e^{-4\Delta^2g^2}/{1+(1+\gamma)e^{-4\Delta^2g^2}}}$.

Now we use the method of Lagrange multipliers to search for the extreme values of $\delta q'$, let $x_1=\cos\theta_1$, $x_2=\cos\theta_2$, $x_3=\cos\phi_0$, $y_1=\sin\theta_1$, $y_2=\sin\theta_2$, $y_3=\sin\phi_0$, we have the Lagrange function
\begin{equation}\begin{split}\label{e73}
F_3=&\frac{2g\Delta^2(1-\gamma)e^{-2\Delta^2g^2}\Delta^2y_1y_2y_3}{1+x_1x_2+(1-\gamma)y_1y_2x_3e^{-2\Delta^2g^2}}+\lambda_1(x_1^2+y_1^2-1)\\
&+\lambda_2(x_2^2+y_2^2-1)+\lambda_3(x_3^2+y_3^2-1).
\end{split}\end{equation}
From $y_1\frac{\partial F_3}{\partial x_1}=x_1\frac{\partial F_3}{\partial y_1}$, we have
\begin{equation}\label{e74}
y_2y_3(x_1+x_2)=0.
\end{equation}
The solution of this equation is $y_2=0$, or $y_3=0$, or
\begin{equation}\label{e75}
x_1=-x_2.
\end{equation}
For $y_2=0$ or $y_3=0$, we have
\begin{equation}\label{e76}
\delta q'=0.
\end{equation}
For $x_1=-x_2$, we have
\begin{equation}\label{e77}
F_3=\pm\frac{2g\Delta^2(1-\gamma)e^{-2\Delta^2g^2}y_3}{1\pm(1-\gamma)x_3e^{-2\Delta^2g^2}}.
\end{equation}
From $y_3\frac{\partial F_3}{\partial x_3}=x_3\frac{\partial F_3}{\partial y_3}$, we have
\begin{equation}\begin{split}\label{e78}
x_3=\mp (1-\gamma)e^{-2\Delta^2g^2},\\
y_3=\pm\sqrt{1-(1-\gamma)^2e^{-4\Delta^2g^2}}.
\end{split}\end{equation}
And the shift
\begin{equation}\label{e79}
\delta q'=\pm\frac{2g\Delta^2(1-\gamma)e^{-2\Delta^2g^2}}{\sqrt{1-(1-\gamma)^2e^{-4\Delta^2g^2}}}.
\end{equation}
From Eqs.(\ref{e76}) and (\ref{e79}), the maximum shift is
\begin{equation}\label{e80}
|\delta q'|_{max}=\frac{2g\Delta^2(1-\gamma)e^{-2\Delta^2g^2}}{\sqrt{1-(1-\gamma)^2e^{-4\Delta^2g^2}}}.
\end{equation}
The maximum shift is obtained on the conditions $\cos\theta_1=-\cos\theta_2$, $\sin\theta_1=\pm\sin\theta_2$, and $\cos\phi_0=\mp(1-\gamma)e^{-2\Delta^2g^2}$.
\subsection{The derivation of Equation (\ref{e34})}
For $|\psi_i\rangle_s=\cos\frac{\theta_1}{2}|0\rangle_s+e^{i\phi_1}\sin\frac{\theta_1}{2}|1\rangle_s$ and $|\psi_f\rangle_s = \cos\frac{\theta_2}{2}|0\rangle_s+e^{i\phi_2}\sin\frac{\theta_2}{2}|1\rangle_s$ are the two vector in Bloch sphere, where $\theta_1\in[0,\pi)$, $\theta_2\in[0,\pi)$, $\phi_1\in[0,2\pi)$, and $\phi_2\in[0,2\pi)$. Eq. (\ref{e33}) could be rewritten as
\begin{equation}\label{81}
\left<\mathbf{O}'\right>=\frac{(1+\cos\theta_1\cos\theta_2-(1-\gamma)\sin\theta_1\sin\theta_2\cos\phi_0)\sin^2g} {1+\cos\theta_1\cos\theta_2+(1-\gamma)\sin\theta_1\sin\theta_2\cos\phi_0\cos2g},
\end{equation}
where $\phi_0=\phi_1-\phi_2\in(-2\pi,2\pi)$. Since $\frac{\partial \left<\mathbf{O}'\right>}{\partial \cos\phi_0}\leq 0$, and $\left<\mathbf{O}'\right>\geq 0$, when $\cos\phi_0=-1$, the reading reaches its maximum value
\begin{equation}\label{e82}
\left<\mathbf{O}'\right>=\frac{(1+\cos\theta_1\cos\theta_2+(1-\gamma)\sin\theta_1\sin\theta_2)\sin^2g} {1+\cos\theta_1\cos\theta_2-(1-\gamma)\sin\theta_1\sin\theta_2\cos2g}.
\end{equation}
Let $t={\cos\frac{\theta_1-\theta_2}{2}}/{\cos\frac{\theta_1+\theta_2}{2}}$, we have
\begin{equation}\label{e83}
\left<\mathbf{O}'\right>=\frac{(1+t^2+(1-\gamma)(1-t^2))\sin^2g}{1+t^2-(1-\gamma)(1-t^2)\cos2g}.
\end{equation}
Since $\frac{\partial \left<\mathbf{O}'\right>}{\partial t^2}\leq 0$, when $t^2=0$, $\left<\mathbf{O}'\right>$ reaches its maximum value
\begin{equation}\label{e84}
\left<\mathbf{O}'\right>_{max}=\frac{(2-\gamma)\sin^2g}{\gamma+2(1-\gamma)\sin^2g}.
\end{equation}
The condition of obtaining maximum value is $\phi_1-\phi_2=\pi$, $\theta_1+\theta_2=\pi$, and the PPS are orthogonal.

\end{document}